\documentclass[twocolumn]{article}
\usepackage{nolta2013}
\usepackage{txfonts}
\usepackage{graphicx}
\begin{document}

\title{A surrogate for networks --- How scale-free is my scale-free network?}

\author{Michael Small, Kevin Judd and Thomas Stemler}
\address{
Department of Mathematics and Statistics, University of  Western Australia\\
Crawley, Western Australia, Australia}

\maketitle

\abstract
Complex networks are now being studied in a wide range of disciplines across science and technology. In this paper we propose a method by which one can probe the properties of experimentally obtained network data. Rather than just measuring properties of a network inferred from data, we aim to ask how typical is that network? What properties of the observed network are typical of all such scale free networks, and which are peculiar?  To do this we propose a series of methods that can be used to generate statistically likely complex networks which are both similar to the observed data and also consistent with an underlying null-hypothesis --- for example a particular degree distribution. There is a direct analogy between the approach we propose here and the surrogate data methods applied to nonlinear time series data.
\endabstract

\section{Introduction}

In surrogate data analysis \cite{jT92} one generates random time series realisations of some underlying null hypothesis. These {\em surrogate} time series are also designed to be largely similar to a set of experimental test data. By comparing statistical properties of the experimental and surrogate data it is possible  to formulate a nonparametric test of the given null hypothesis.   A simple example of this type of process would be random shuffling of the order of the experimental data: by comparing an experimental data set to randomly shuffled surrogates we test whether there is any temporal structure in the experimental data.

There have been some recent attempts to extend the idea of surrogate data analysis to complex network (for example \cite{dB13}). Surrogate data methods are applied to time series as a sort of ``sanity-check'' before one proceeds to more complex nonlinear methods \cite{casisurr}. When studying complex networks, there is often a need for a similar procedure --- loosely speaking, it may be necessary to ask whether an experimentally obtained complex network is ``non-trivial''. Analogously to the simple time series shuffling method, the most obvious way to do this is to shuffle the elements of the adjacency matrix (while preserving symmetry).  Doing this will preserve the degree distribution of the network, but otherwise generate a random graph. This is fine for comparing the observed network properties (diameter, assortativity, clustering and so-on \cite{demnet,nolta2012}) to an ensemble of random networks with an identical histogram of node degrees\footnote{In what follows we make a distinction between the empirical discrete histogram of node degrees for a given graph, and the idealised degree distribution representative of the asymptotic (in both network size and sample mean) behaviour of a family of such graphs.}. However, it is neither clear what this ensemble is representative of, nor how to generalise to more complicated sampling of families of graphs.

We have recently introduced a new method \cite{kJ13} to randomly sample graphs with a given degree distribution --- where all {\em connected}\footnote{Connectivity is an additional constraint we choose to add to the scheme. We do this because, to us, connected graphs are the interesting ones. Nonetheless, it is trivial to remove this constraint or impose other desirable features.} graphs with a given histogram are considered to be equally likely. A maximum likelihood expression is used to evaluate the probability of realising a particular histogram given the desired degree distribution. Monte Carlo Markov Chain (MCMC) sampling is used to ensure proper sampling of the likelihood function. 

Using this technique we found several interesting properties of scale-free (that is, power-law degree distribution) graphs which are not widely acknowledged. Moreover, we extended the method to generalise the preferential attachment growth model \cite{prefpa} and showed that the maximum likelihood approach to growing a scale-free network actually deviates significantly from the accepted Barab\'asi-Albert model \cite{aB99} (BA). However, it is not these properties that concern us here.

\section{Method}

In this paper we apply the same methodology to generate random realisations from a specified degree distribution starting with an experimentally obtained seed network. That is, we take an experimental network which we wish to test and then generate multiple random samples from the family of all networks consistent with a particular degree distribution. Since theses random realisation start from the observed experimental network they are {\em constrained} realisation in the same way that we usually aim to generate constrained surrogate data sets. Nonetheless, these random network are also realisations from the family of all networks consistent with a particular degree distribution\footnote{Any other particular desirable traits could also be imposed as constraints on the surrogate network generation process.}. Hence, by comparing statistical properties of the surrogate networks to the experimental network we are able to test whether a particular experimental network is actually typical of that particular distribution. Moreover, in cases where the statistical properties differ we can also say something --- based on which statistics diverge --- about how the particular network is atypical.  

The method we propose in \cite{kJ13} is a direct implementation of the MCMC technique and the maximum likelihood optimisation of a given degree distribution. Nonetheless, it can be computationally rather burdensome. To extend the technique to more practical networks, we make several simplifications. The technique will be described in more detail in future, nonetheless, the algorithm is described briefly as follows.

Let $A_{ij}$ denote the $(i,j)-$th element of the adjacency matrix of an $N$-node network. That is,
\begin{eqnarray}
A_{ij} &=&\left\{\begin{array}{cc}
1 & \mbox{if node-$i$ and node-$j$ are linked}\\
0 &\mbox{otherwise}
\end{array}\right.
\end{eqnarray}
such that  $A_{ii}=0$ and $A^{T}=A$.
Let $t(x)$ denote the desired cumulative distribution ($t(x)\equiv p(x)$ from above)  and $\hat{p}(x)$ the sample cumulative distribution. That is,
\begin{eqnarray}
t(x)&=&{\rm Prob}(\#({\rm node}) \leq x)\\
\hat{p}(x)&=&\frac{1}{N}\sum_{i=1}^N H(\#(\mbox{node-$i$})\leq n)
\end{eqnarray}
 for a network of $N$ nodes, where $\#({\rm node})$ denotes the degree of the node and $H(X)=1$ iff $X$ is true ($H(X)=0$ otherwise). The Kolomogorov-Smirnov test statistic is given by $\max_x|h(x)|$ where
 \begin{eqnarray}
 h(x) & = t(x)-\hat{p}(x).
 \end{eqnarray}
Let $h^t(x)$ denote the value of $h(x)$ after $t$ steps of the following algorithm and define
\begin{eqnarray}
m & = & \max_x h(x)\\
w & = & \arg\max_x h(x)\\
n & = & \min_x h(x)\\
u & = & \arg\min_x h(x)  
\end{eqnarray}
Hence, $w$ is the degree which is most under-represented in the sample distribution $\hat{p}(x)$ and $u$ is the degree which is most over-represented. Here, and in all of what follows, we assume that $m>0>n$ --- this condition can be guaranteed by ensuring that the average degree of $A$ matches the expected degree of the target distribution
\begin{eqnarray}
\label{samemean} \frac{1}{N}\sum_i\sum_jA_{ij} &=& \int_xxt'(x)dx\\
&\approx& \sum_{x=1} x(t(x)-t(x-1))
\end{eqnarray}
depending on whether $x(t)$ is treated as discrete or continuous.

At each step, the new distribution $h^{t+1}(x)$ is obtained from $h^t(x)$ by modifying a node of degree $u$ to make it into a node of degree $w$. By so doing $m\rightarrow m-1$ and $n\rightarrow n+1$. By construction, a sequence of such moves will cause $\max_x|h^t(x)|\rightarrow 0$  as $t\rightarrow\infty$ --- but only provided the additional or deleted links are chosen in such a way as to also guarantee that the $h^t(x)\rightarrow 0$ point-wise. Hence, we require that each of these moves does not make the distribution worse by adversely choosing the neighbors (new of old neighbours) of the candidate node-$i$. 

That candidate node-$i$ is chosen such that $\#(\mbox{node-$i$})=u$. If $u>w$ then delete (existing links) or disconnect (and reconnect elsewhere)  $(u-w)$ links from node-$i$. Otherwise, $u<w$ and add (new links) or reconnect (existing links, disconnected from other nodes) $(w-u)$ links. Let $j\neq i$ denote the index of a random node, node-$j$. Note that if $h(\#({\rm node-}j))>0$ then the current distribution $\hat{p}(x)$  has insufficient nodes of degree $\#({\rm node-}j)$. Conversely, if $h(\#({\rm node-}j))<0$ then there are two many nodes of degree $\#({\rm node-}j)$. For the sake of conciseness, let $h_{\#}(j):=h(\#({\rm node-}j))$.

We now need to make $|w-u|$ modifications (either additional links or removing links) to the degree of node-$i$, each modification will follow the follow scheme:
\begin{enumerate}
\item If $w>u$ (we want to increase the degree of node-$i$ from $u$ to $w$). Choose $j\neq i$ such that $A_{ij}=0$ (find a candidate non-neighbour) 
\begin{enumerate}
\item \label{up-more} If $h_{\#}(j)\geq 0$ (we need more nodes like node-$j$ and so should preserve its degree) choose $k$ such that $A_{jk}=1$ and $h_{\#}(k)$ is minimal. That is, 
\begin{eqnarray}
k &=& \arg\min_{\tiny \begin{array}{c} k\\A_{jk}=1,A_{ki}=0\end{array}}h(\#({\rm node-}k)).
\end{eqnarray}
To ensure that $h^t(x)\rightarrow 0$ pointwise, we need that $h_\#(k)<0$. Disconnect node-$j$ from node-$k$ and connect it to node-$i$ instead: set 
$A_{jk}=A_{kj}=0$ and $A_{ij}=A_{ji}=1$

Note that, rather than choose $k$ such that $h_{\#}(k)$ is minimal, one could also introduce a preferential mechanism here --- choose node-$k$ with probability 
\begin{eqnarray}
q(k)&\propto &\left\{\begin{array}{cc}-h_{\#}(k) &\mbox{if $h_{\#}(k)<0$}\\0 &\mbox{otherwise}\end{array}\right..
\end{eqnarray}

\item \label{up-less} If $h_{\#}(j)<0$ (we've already got too many nodes like this one --- no harm to have one less) then add a new link from node-$j$ to node-$i$: set $A_{ij}=A_{ji}=1$.
\end{enumerate}
\item Otherwise, $w<u$ (we want to decrease the degree of node-$i$ from $u$ to $w$). Choose $j\neq i$ such that $A_{ij}=1$ (find a candidate neighbour to delete) 
\begin{enumerate}
\item  \label{down-more} If $h_{\#}(j)\geq 0$ (again, nodes like node-$j$ are under-represented and hence should be preserved) choose $k$ such that $A_{jk}=0$ and $h_{\#}(k)$ is minimal:
\begin{eqnarray}
k &=& \arg\min_{\tiny \begin{array}{c}k\\A_{jk}=1\end{array}}h(\#({\rm node}-k)).
\end{eqnarray}
Disconnect node-$j$ from node-$i$ and connect node-$j$ to node-$k$ instead: set $A_{jk}=A_{kj}=1$ and $A_{ij}=A_{ji}=0$

As before, poointwise convergence of $h^t(x)$ requires that $h_\#(k)<0$. Again, we could introduce the equivalent preferential attachment mechanism here. Rather than choose $k$ such that $h_{\#}(k)$ is minimal, choose node-$k$ with probability 
\begin{eqnarray}
q(k)&\propto &\left\{\begin{array}{cc}-h_{\#}(k) &\mbox{if $h_{\#}(k)<0$}\\0 &\mbox{otherwise}\end{array}\right..
\end{eqnarray}

\item  \label{down-less} If $h_{\#}(j)<0$ (we've already got too many nodes like this one --- no harm to have one less) then delete the link from node-$j$ to node-$i$: set $A_{ij}=A_{ji}=0$.
\end{enumerate}
\end{enumerate}

Note that the substitution described in the operations  (\ref{up-more}) and (\ref{down-more}) increase or decrease (respectively) the degree of node-$i$ by rewiring the links of node-$j$ --- the degree of a third party, node-$k$, is altered. To ensure the point-wise convergence of $h^t(x)\rightarrow 0$ we can stimulate that $h_\#(k)<0$. In general this may not be necessary -  a slight elevation from uniform improvement of $h^t(x)$ at each step may still yield a practical solution. Nonetheless, the overall number of links is not changed by the operations  (\ref{up-more}) or (\ref{down-more}), and the third party node-$k$ is chosen to be the most amenable donor or receiver (respectively) of the link with node-$i$. In contrast, the operations described by  (\ref{up-less}) and (\ref{down-less}) do change the overall number of links --- step (\ref{up-less}) adds new links while step (\ref{down-more}) deletes existing connection. Assume that the distribution $h^t(x)$ satisfies $\left<h^t(x)\right>_x=0$ (see Eqn. \ref{samemean}), then these operations should appear with equal probability. 

Finally, after each modification of $A$ which results in the deletion of at least on link, we need to check that the resulting network remains connected (i.e. the giant component is preserved). Note that (as we did before \cite{kJ13}) a network is connected if the matrix $C_p=I+A+A^2+A^3+\ldots+A^p$ has no zero elements ($p=N-1$). 
Two nodes, $i$ and $j$ are connected if there exists $p<N$ such that $C_{p,ij}\neq 0$. It is therefore necessary for us to check this condition each time we perform an edge deletion (in operation (\ref{up-more}) it is $A_{jk}=0$ and in step (\ref{down-more}) and (\ref{down-less}) it is $A_{ij}=0$ --- step (\ref{up-less}) involves no edge deletion and checking is therefore not required). This can be computed efficiently as follows. If $a_i$ and $a_j$ are columns $i$ and $j$ of $A$ and $w_{j,p+1}=Aw_{j,p}+w_{j,p}$ $w_{j,1}=a_j$ then $C_{p,{ij}}=a_i^tw_{j,p}$. If an operation would result in a non-connected component, then that operation is abandoned (and perhaps another is tried instead).

This algorithm does not have a clearly defined stopping criterion --- unless $h^t(x)=0$ is achieved. In all other cases practical constraints will dictate how long to continue. In exactly the same spirit as the popular iterated amplitude adjusted Fourier transform surrogate technique \cite{tS96} one continues for long enough to converge on the desired distribution but not so long as to eliminate all variability.

\begin{figure}
\begin{center}
\includegraphics[width=0.45\textwidth]{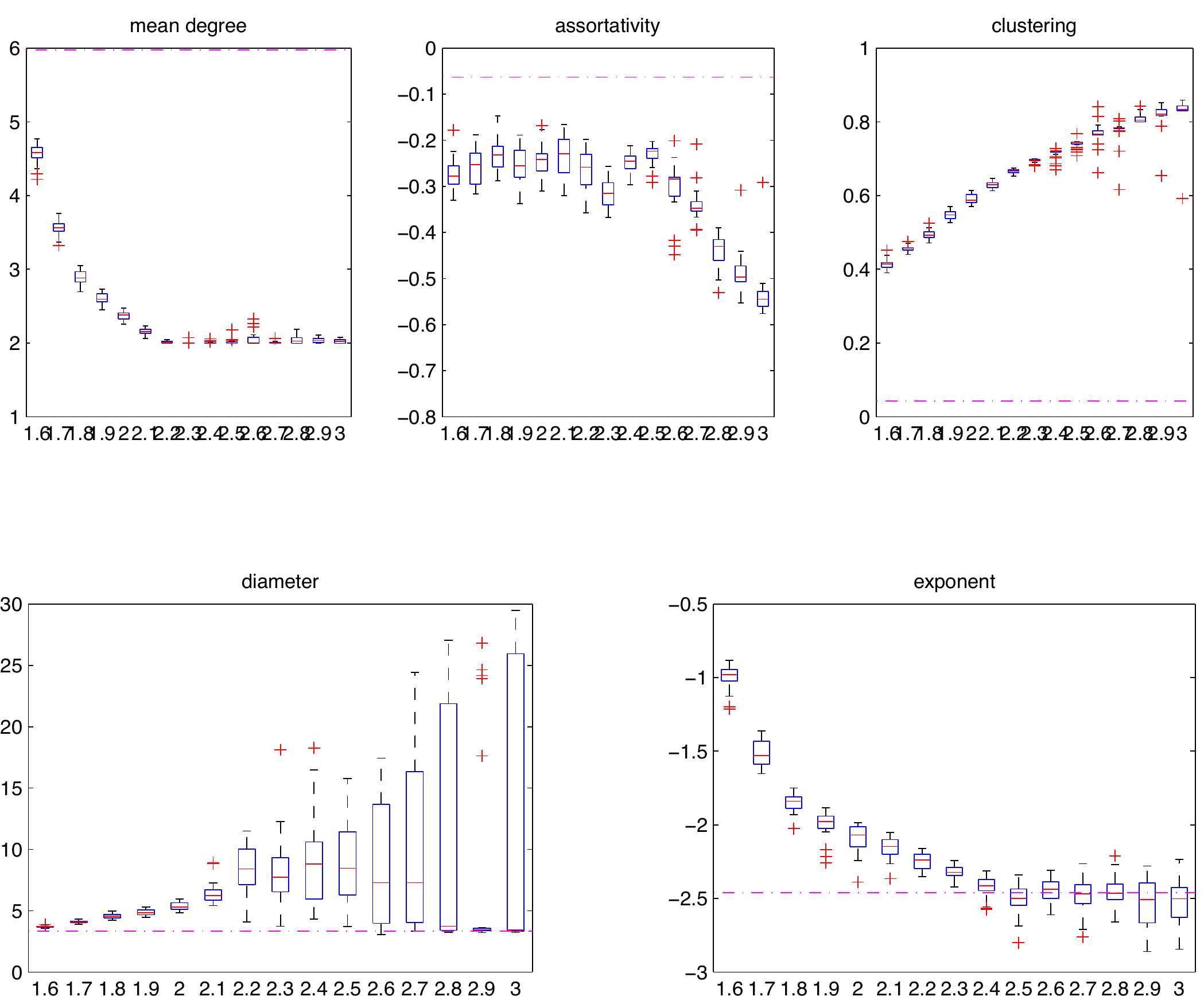}
\end{center}
\caption{{\bf Reject.} The five panels depict estimates of the five most usual complex network characteristics: mean degree; assortativity; clustering; network diameter (mean path-length); and, degree distribution exponent ($\gamma$). Each panel depicts the value estimated from the original data (magenta dot-dashed line) and box plot distributions (25-th to 75-th percentile as a box; whiskers to range of the data, excluding algorithmically identified outliers; and, outliers plotted as individual crosses) as a function of the chosen value of  $\gamma\in[1.6,3]$ specified for the surrogate generation process. In each case, the (same) original network was constructed with BA preferential attachment with $500$ nodes and $m=3$ nodes added at each time step. The null hypothesis is that the degree distribution conforms to $\frac{1}{\zeta{(\gamma)}} k^{-\gamma}$ for $k\geq1$. For each value of $\gamma\in[1.6,3]$, $30$ surrogate networks are randomly generated with the algorithm described in the text. Mean degree, assortativity and clustering all provide evidence with which to reject the null.}
\label{fig1}
\end{figure} 

\begin{figure}
\begin{center}
\includegraphics[width=0.45\textwidth]{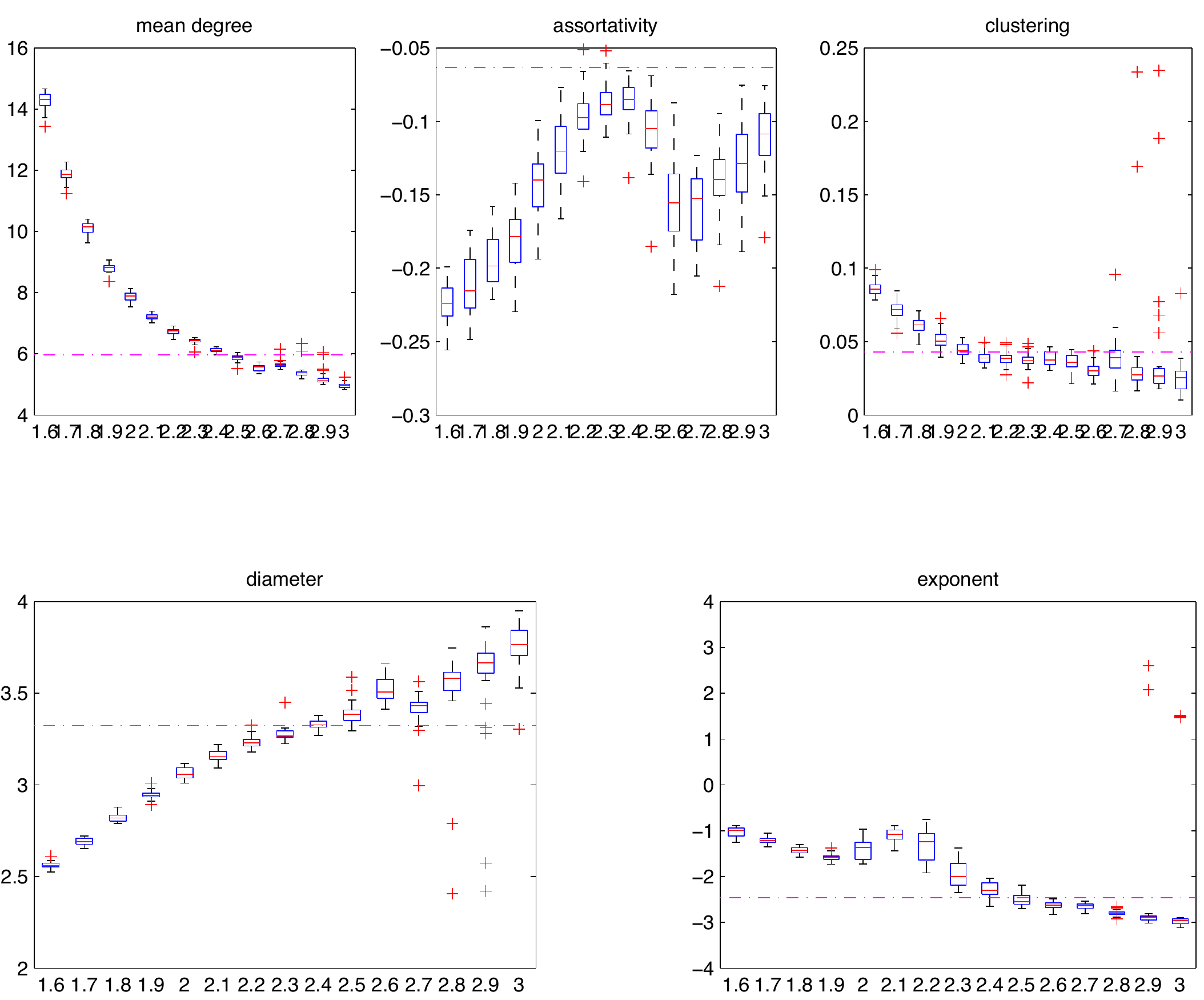}
\end{center}
\caption{{\bf Accept.} The calculation of Fig. \ref{fig1} is repeated, with the additional constraint on the surrogates that the minimum degree $m=3$. That is, the hypothesis being tested here is that the degree distribution is  proportional to $k^{-\gamma}$ for $k\geq 4$ and $0$ otherwise. This matches the BA preferential attachment algorithm and hence, for the correct range of degree distribution exponent $\gamma$ we are unable to reject the null.}
\label{fig2}
\end{figure} 

\section{Results}

In this section we present a very simple test of our method. We generate a $500$ node network using the BA preferential attachment algorithm --- adding $m=3$ new links with each new node. This, asymptotically, generates a network with degree distribution$p(k)$ given by 
\begin{eqnarray}
p(k)=\left\{
\begin{array}{cc}
\frac{1}{\zeta(\gamma)}(k-m+1)^{-\gamma} & k\geq m\\
0 & k < m
\end{array}\right.
\label{eq1}
\end{eqnarray}
From this single random network we then generate surrogate networks with the scheme described in the proceeding section: first, imposing no minimum degree (i.e. $m=1$); and, second, imposing the correct minimum degree $m=3$. The surrogate networks are constructed to conform to a specified degree distribution, and so we repeat the procedure for $\gamma\in[1.6,3]$. In each case, we compute the usual assortment of complex network statistics (see Fig. \ref{fig1}, caption) \cite{mN10b} and compare that distribution to the value for the original network. By doing so, we can directly deduce from the results in Fig. \ref{fig1} that the 
BA preferential attachment is not consistent with a scale-free distribution with no minimum degree --- the networks have statistically significant difference. However, as expected, the network is consistent with shifted power-law with $m=3$ (\ref{eq1}).

\section{Conclusion}

In this communication we have presented a prototype of a method which allows one to test whether a given set of experimental network data is consistent with a specified hypothesis. That hypothesis is stated in terms of the expected degree distribution and using an MCMC algorithm random realisations consistent with both that hypothesis and yet still {\em like} the original network data are generated. This approach is an exact analogy to the method of surrogate data. 

We demonstrate that the method is able to correctly identify the underlying model responsible for specific network data --- even in the case of a rather small network.

\section*{Acknowledgement}

MS is support by an Australian Research Council Future Fellowship (FT110100896).

%\bibliography{../../manuscripts/bibliography}
%\bibliographystyle{abbrv}

\end{document}